%% file: main.tex
\documentclass[ALICE,manyauthors]{cernphprep}
\usepackage[comma,square,numbers,sort&compress]{natbib}
\usepackage[T1]{fontenc} 
\usepackage[utf8]{inputenc}
\usepackage[english]{babel}
\usepackage{amssymb}
\usepackage{hyperref}
\usepackage{lineno}
\usepackage[amssymb]{SIunits}
\usepackage{booktabs}
\usepackage{multicol}
\usepackage{multirow}
\usepackage{xcolor}
\usepackage{overpic}
\usepackage{units}
\usepackage{placeins}
\usepackage{graphics}
\usepackage{textcomp} 
\usepackage{amsmath}
\usepackage{soul}
\usepackage{orcidlink}

\newcommand{\cms}          {\ensuremath{\sqrt{s}}}

\newcommand{\dndeta}       {\ensuremath{\mathrm{d}N_\mathrm{ch}/\mathrm{d}\eta}}
\newcommand{\dNdeta}       {\ensuremath{\langle\dndeta\rangle}}
\newcommand{\pT}           {\ensuremath{p_\mathrm{T}}}
\newcommand{\pTcut}        {\ensuremath{p_\mathrm{T}}^\mathrm{cut}}

\newcommand{\avdndeta}{\langle \mathrm{d}N_\mathrm{ch} / \mathrm{d}\eta \rangle}

\newcommand{\inelg}{\mathrm{INEL}\text{>}0}
\newcommand{\inelf}{\inelg_{p_{\mathrm{T}}>0.15}^{|\eta|<0.8}}
\newcommand{\inels}{\inelg_{p_{\mathrm{T}}>0.5}^{|\eta|<0.8}}
\newcommand{\inelt}{\inelg_{p_{\mathrm{T}}>1}^{|\eta|<0.8}}
\newcommand{\inelq}{\inelg_{p_{\mathrm{T}}>2}^{|\eta|<0.8}}

\newcommand{\inelss}{\inelg_{p_{\mathrm{T}}>0.5}^{|\eta|<2.4}}
\newcommand{\inelsss}{\inelg_{p_{\mathrm{T}}>0.5}^{|\eta|<2.5}}

\begin{document}%
\PHyear{2022}       
\PHnumber{262}      
\PHdate{18 November}  
\begin{titlepage}
 \title{Pseudorapidity densities of charged particles with transverse momentum thresholds in pp collisions at $\sqrt{\mathbf{\textit{s}}}=$ 5.02 and \unit[13]{TeV}}
\ShortTitle{Pseudorapidity densities  with minimum-$\pT$ thresholds }  

\Collaboration{ALICE Collaboration\thanks{See Appendix~\ref{app:collab} for the list of collaboration members}}
\ShortAuthor{ALICE Collaboration} 

\begin{abstract}

The pseudorapidity density of charged particles with minimum transverse momentum ($\pT$) thresholds of 0.15, 0.5, 1, and \unit[2]{GeV$/c$} is measured in pp collisions at the centre of mass energies of $\cms =$ 5.02 and \unit[13]{TeV} with the ALICE detector. The study is carried out for inelastic collisions with at least one primary charged particle having a pseudorapidity ($\eta$) within $\pm0.8$ and $\pT$ larger than the corresponding threshold. In addition, measurements without $\pT$-thresholds are performed for inelastic and non-single-diffractive events as well as for inelastic events with at least one charged particle having $|\eta|<1$ in pp collisions at $\cms =$ 5.02 TeV for the first time at the LHC. These measurements are compared to the PYTHIA 6, PYTHIA 8, and EPOS-LHC models. In general, the models describe the $\eta$ dependence of particle production well. However, discrepancies are observed for the highest transverse momentum threshold ($\pT>2 {\rm\ GeV}/c$), highlighting the importance of such measurements for tuning event generators. The new measurements agree within uncertainties with results from the ATLAS and CMS experiments obtained at $\cms = 13$ TeV. 
\end{abstract}
\end{titlepage}

\section{Introduction}

The pseudorapidity density of charged particles, \dndeta, is a key observable for understanding the general properties of particle production in high-energy hadronic collisions. 
At collider energies, particle production in proton--proton (pp) collisions has origins in both soft and hard processes~\cite{Field:1976ve}. Hard processes are those with high enough transverse momentum transfer  ($Q\gg \Lambda_{\rm QCD}\sim 200 \rm{\ MeV}$) between the scattering partons such that they can be described by perturbative quantum chromodynamics (pQCD)~\cite{Srednicki:2007qs}. For the description of soft processes, non-perturbative phenomenological models inspired by pQCD and implemented in modern Monte Carlo (MC) generators are needed~\cite{Collins:1977jy,Greiner:2002ui,Ostapchenko:2007qb,Fletcher:1994bd,Engel:1994vs,Sjostrand:2006za,Pierog:2013ria,Bahr:2008pv,Gleisberg:2008ta}.
The measurement of the charged particle pseudorapidity density provides constraints on the descriptions of particle production mechanisms and input for tuning of MC event generators, such as PYTHIA and EPOS used for physics at hadron colliders~\cite{Sjostrand:2006za,Skands:2010ak,Sjostrand:2007gs,Sjostrand:2014zea,Skands:2014pea,Pierog:2013ria}.

Following earlier ALICE studies of particle production in pp collisions~\cite{Aamodt:2009aa, Aamodt:2010ft, Aamodt:2010pp, ALICE:2015olq, Adam:2015pza, ALICE:2020swj}, this publication presents a set of measurements of the pseudorapidity density of primary charged particles in inelastic events (INEL), non-single-diffractive events (NSD), and inelastic events with at least one charged particle in $|\eta|<1$ ($\inelg$) for pp collisions at $\sqrt{s} =$5.02 TeV. In ALICE, primary charged particles are defined as charged particles with a mean proper lifetime $\tau$ larger than 1 cm/$c$, which were produced either promptly at the primary vertex or from decays of particles with $\tau<1$ cm/$c$ restricted to decay chains leading to the interaction~\cite{ALICE-PUBLIC-2017-005}. 
In the previous measurements of Refs.~\cite{Aamodt:2009aa, Aamodt:2010ft, Aamodt:2010pp, ALICE:2015olq, Adam:2015pza, ALICE:2020swj}, $\dndeta$ was reported for the INEL, NSD, and $\inelg$ event classes and without any selection on the transverse momentum (\pT) of the particles, i.e., for $p_{\rm T}>0$.

In order to obtain improved constraints on models of charged particle production in hard processes, the study is extended to measurements of pseudorapidity densities of primary charged particles with transverse momenta $\pT>\pTcut$, where $\pTcut$ $= 0.15, 0.5, 1$, or 2 GeV/$c$, for different event classes with at least one charged particle in $|\eta|<0.8$ with a $\pT$ larger than the corresponding threshold $\pTcut$ at $\sqrt{s} =$ 5.02 and \unit[13]{TeV}. The four event classes associated with the different $\pT$ thresholds are identified as $\inelf$, $\inels$, $\inelt$, and $\inelq$. These measurements are an extension of the previous studies at LHC Run 1 collision energies ($\sqrt{s} =$ 0.9 and \unit[7]{TeV})~\cite{ALICE:2013bva}. 
The $\pT$ threshold of 0.15 GeV/$c$ is chosen to allow comparisons to ALICE results at lower $\sqrt{s}$ while the 0.5 GeV/$c$ threshold allows comparisons to ATLAS and CMS results~\cite{Aad:2010ac,Aad:2016mok,CMS:2018nhd}. The higher thresholds of 1 and 2 GeV/$c$ enable the study of particle production with harder particles.

The article is organised as follows. Section 2 addresses the experimental conditions and data samples used in the analysis. Then, the analysis procedures to measure the primary charged particle production and the applied corrections are explained in Sec.~3. Section~4 describes the systematic uncertainties while the results compared to those of ATLAS and CMS and to model predictions are presented in Sec.~5. A brief summary and conclusions are given in Sec.~6.

\section{Experimental conditions and data collection}
The data samples used in this analysis were collected during LHC Run 2. A sample of $4.4\times10^6$ minimum bias events in pp collisions at $\sqrt{s} = 5.02$ TeV was used for measurements without a $\pT$ threshold. For measurements requiring a minimum track $\pT$, samples of $2.5\times10^8$ and $2.8\times10^7$ minimum bias events in pp collisions at $\sqrt{s} = 5.02$ and 13 TeV, respectively, were analysed.

Detailed information about the ALICE detector and its performance during LHC Run 2 can be found in Refs.~\cite{Aamodt:2008zz} and~\cite{ALICE:2014sbx}. 
Tracking of charged particles is mainly performed with the Inner Tracking System (ITS)~\cite{aliceITS} and Time Projection Chamber (TPC)~\cite{ALICE:2014sbx} located inside a large solenoid that produces a homogeneous magnetic field of 0.5 T directed along the beam direction ($z$ axis in the ALICE reference frame). 

The detector closest to the interaction point is the ITS which is composed of 6 cylindrical layers of high resolution silicon detectors. The innermost two layers consist of the Silicon Pixel Detector (SPD)~\cite{Santoro:2009zza,Abelev:2014ffa}. The SPD layers are coaxial to the beam line with radii of 3.9 and \unit[7.6]{cm} covering the pseudorapidity range $|\eta|<2$ for the first layer and $|\eta|<1.4$ for the second layer. An enlarged pseudorapidity coverage of $|\eta|<2$ is reached using events whose primary vertex along the beam direction ($z_\mathrm{vtx}$) is within \unit[$\pm10$]{cm} from the nominal interaction point ($z_\mathrm{vtx}=0$). Counting the number of tracks for analysis without a $\pT$ threshold (INEL, NSD, $\inelg$) relies on the reconstruction of tracklets, which are track segments connecting hits on the two SPD layers and pointing to the primary vertex. Due to the bending of particle trajectories in the magnetic field and multiple scattering, the reconstruction efficiency limits their measurements to $\pT > 50$ MeV/$c$~\cite{ALICE:2015olq}. 

The TPC~\cite{ALICE:2014sbx}, located outside the ITS, is a \unit[90]{$\mathrm{m^3}$} cylindrical drift chamber. The TPC covers the pseudorapidity range $|\eta|<0.9$ with respect to $z=0$ and the full azimuthal angle. It provides excellent momentum and spatial resolutions for tracking of charged particles. The V0 detector~\cite{forwarddetectorsTdr} consists of two scintillator arrays that are located on each side of the interaction point along the beam direction and cover the pseudorapidity regions $-3.7 < \eta < -1.7$ (V0C) and $2.8 < \eta < 5.1$ (V0A). It is used for triggering and event selection. 

To select different event classes, the SPD and the V0 detectors are used. For the measurements of \dndeta\ in INEL, NSD, and $\inelg$ events, the minimum-bias trigger requires a hit in the SPD or in either one of the V0 arrays. For the analyses requiring a minimum $\pT$ threshold, the minimum-bias trigger requires signals on both sides of the V0.
The SPD and V0 detectors are also used to suppress background from beam--gas collisions and other machine-induced backgrounds. The contamination from background events is removed offline by using the timing difference between the signals in the V0A and V0C detectors~\cite{ALICE:2014sbx}, exploiting the V0 time resolution that is better than 1 ns.
Background events are also rejected by exploiting the correlation between the number of clusters on both layers of the SPD and the number of tracklets in the SPD.

Another type of event background comes from pileup, happening when multiple collisions occur in the same bunch crossing. The overall probability of pileup in ALICE is around $10^{-3}$ in the minimum-bias pp samples used for these analyses~\cite{ALICE:2015olq}. Pileup contamination is reduced by rejecting events with multiple interaction vertices reconstructed from SPD tracklets. The remaining undetected pileup is negligible in the data samples considered for the analysis presented in this article. 

The position of the interaction vertex is obtained using two different approaches: the first is based on the hits in the two SPD layers, the second utilises global tracks that are reconstructed in the TPC and matched to ITS clusters~\cite{ALICE:2014sbx}. The primary vertex position is required to be in $|z_\mathrm{vtx}|<10$ cm for both inclusive and $\pT$ threshold \dndeta\ studies.

\section{Analysis procedure and corrections}

The measurements of $\dndeta$ in the event classes without a $\pT$ threshold (INEL, NSD, $\inelg$) are based on the tracklet counting method which was used for previous inclusive $\dndeta$ measurements~\cite{Aamodt:2009aa,Aamodt:2010ft,Aamodt:2010pp,ALICE:2015olq,Adam:2015pza,ALICE:2020swj}.
For SPD tracklets, the association to the position of the primary vertex of the collision is ensured through a $\chi^2$ requirement. By using the interaction point reconstructed with the SPD as the origin, differences in the azimuthal ($\Delta\varphi$, bending plane) and polar ($\Delta\theta$, non-bending direction) angles of two hits, one in the inner and one in the outer SPD layer, are calculated. The tracklets are selected with the following quality cut
\begin{equation}
\label{chi2equation}
	\chi^2 = \frac {(\Delta \varphi)^2}  {\sigma ^ 2 _\varphi}  +  \frac{1 }{ \sin^2 \big(\frac{\theta_1+\theta_2}{2}\big)}    \times  \frac  {(\Delta \theta)^2}   {\sigma^2_{\theta}} < 1.6,
\end{equation}
where $\sigma_\varphi$ = 0.08 rad, $\sigma_\theta$ = 0.025 rad, and $\theta_1$ and $\theta_2$ are the polar angles of the hits in each layer of the SPD~\cite{ALICE:2015olq}.

To select primary charged particles for the results with $\pT$ thresholds, tracks reconstructed using the hits in the ITS and TPC (global tracks)~\cite{Aamodt:2008zz} are allowed for counting and momentum measurements of charged particles in ALICE. High-quality tracks are selected by requiring them to have at least 70 (out of maximally 159) crossed pad rows in the TPC, have a good quality of the track momentum fit ($\chi^2/\mathrm{ndf}<2$), have a distance of closest approach to the primary vertex along the $z$ direction ($\mathrm{DCA}_{z}$) lower than 2 cm, and have a transverse DCA ($\mathrm{DCA}_{xy}$) lower than $0.0105+0.035 p_\mathrm{T}^{-1.1}$ cm (7 times larger than its resolution) with $\pT$ in units of GeV/$c$~\cite{ALICE:2017ban}. 

The primary vertex using the SPD is reconstructed by correlating hits in the two SPD layers. The resolution of the SPD vertex is on average \unit[30]{\textmu m}~\cite{ALICE:2015olq}.
The primary vertex reconstructed using the ITS and the TPC is called the global track vertex. For global track vertices, the resolution is typically \unit[100]{\textmu m} in the longitudinal ($z$) and \unit[50]{\textmu m} in the transverse ($xy$) direction. Both SPD and global track vertices must be present and consistent by requiring that the difference between the two $z$ positions is less than 5 mm.

All corrections are calculated using MC events generated with PYTHIA 6 with the Perugia 2011 tuning~\cite{Sjostrand:2006za,Skands:2010ak} or PYTHIA 8 with the Monash 2013 tuning~\cite{Sjostrand:2006za,Sjostrand:2007gs,Sjostrand:2014zea,Skands:2014pea} event generators with particle transport performed via a GEANT3~\cite{Brun:1994aa} simulation of the ALICE detector.
Three different MC corrections are applied to the raw $\dndeta$: 
(a) a track-to-particle correction that accounts for detector inefficiencies and background particles like secondaries from interactions in the detector material and decays of primary charged particles, including the strange particle content correction, (b) a vertex reconstruction efficiency correction for triggered events without a reconstructed vertex, and (c) a trigger efficiency correction, which accounts for the bias due to the trigger requirement for the corresponding event class. All the track-to-particle corrections are applied as a function of $z_\mathrm{vtx}$ in order to consider $z$-dependent $\dndeta$ efficiency. The values for the three event classes (INEL, NSD, INEL>0) at $\eta=0$ for the track-to-particle correction are $\sim$(1.34, 1.34, 1.33), the vertex reconstruction efficiency correction is $\sim$(0.89, 0.94, 1.0), and the trigger efficiency is $\sim$(0.88, 0.96, 0.97), respectively.
 
The ALICE definition of primary charged particles excludes particles originating from weak decays of strange particles. Therefore, data have to be corrected for cases when daughter particles from these decays pass the track selection. The strangeness content in data is 40--50\% larger than in PYTHIA 6 and PYTHIA 8~\cite{ALICE:2016fzo} in both inclusive and $\pT$-threshold analyses. This discrepancy is accounted for by scaling the strangeness content in the MC simulation to that in the data. The scaling factor is determined by measuring the ratio between the abundance of decay products in $|\eta|<0.8$ from reconstructed $\mathrm{K}^0_\mathrm{s}$, $\Lambda$, and $\bar{\Lambda}$ in data and MC simulations. The scaling factor in the correction procedure compensates for the underestimated strangeness content in MC and result in a relative downward correction of about 0.5\% on the final $\dndeta$.

The results of \dndeta\ for INEL and NSD events are affected by the model uncertainty for diffractive events. Cross-section measurements with ALICE indicate that the number of single-diffractive (SD) and double-diffractive (DD) events are about 20\% and 12\% of the number of inelastic events, respectively, at both $\sqrt{s} = 2.76$ and 7 TeV~\cite{Abelev:2012sea}. 
As for the previous ALICE measurements of $\dndeta$ for the INEL, NSD, and $\inelg$ event classes~\cite{ALICE:2015olq}, a special PYTHIA 6 tune for diffraction is used~\cite{Abelev:2012sea}. For this tune, the diffractive mass distribution of SD events is re-weighted while the one of DD events is unchanged. For a consistent treatment, the mass distribution of SD events in both event generators (PYTHIA 6 Perugia 2011 and PYTHIA 8 Monash 2013) are re-weighted to follow those used in the special PYTHIA 6 tune. Note that both the $\inelg$ and $\pTcut$ analyses do not contain single diffractive events due to the requirement of at least one charged particle at midrapidity.  Therefore, the tuning procedure for the SD diffraction mass in the event generator is not required.

\section{Systematic uncertainties}

\begin{table}[htbp]
\caption{Relative values of systematic uncertainties (expressed in \%) on $\dndeta$ at $\eta=0$ for INEL, NSD, and $\inelg$ event classes determined in pp collisions at $\cms = 5.02$ TeV. }
\centering
\begin{tabular}{cccc}
\hline\hline
\multirow{2}{*}{Source of uncertainty} &   \multicolumn{3}{c}{ Systematic uncertainty at $\eta=0$ ($\%$)} \\ 
&  INEL &  NSD & $\inelg$ \\
\hline\hline
Diffraction ratio &  $\pm4.5$ & $\pm2$ & $\pm0.1$\\
\hline
Diffraction shape  & $+3$ & $-0.2$ & $-0.2$ \\
\hline
Zero-\pT\ extrapolation & $+1, -0.5$ &  $+1, -0.5$ &  $+1, -0.5$\\
\hline
Event generator dependence & $\pm0.2$ & $\pm1$ & $\pm0.4$\\
\hline
Acceptance and efficiency & $\pm0.8$ & $\pm0.8$ & $\pm0.8$ \\
\hline
$z_{\mathrm{vtx}}$ dependence &  $\pm0.3$ & $\pm0.3$ & $\pm0.3$\\
\hline
Strangeness enhancement factor & $\pm0.5$ &  $\pm 0.5$ & $\pm 0.5$\\
\hline
Particle composition & $\pm0.4$ & $\pm 0.4$ & $\pm 0.4$ \\
\hline
Material budget & $\pm0.2$ & $\pm0.2$ & $\pm0.2$\\
\hline\hline
Total systematic uncertainty& $+5.5, -4.6$ & $+2.5, -2.3$ & +1.2, -1.1\\
\hline\hline
\end{tabular}

\label{tab:dndetasystracklets}
\end{table}%

\begin{table}[htbp]
\centering
\caption{Relative values of systematic uncertainties (expressed in \%) on $\dndeta$ at $\eta=0$ for $\inelf$, $\inels$, $\inelt$, and $\inelq$ event classes determined in pp collisions at $\cms = 5.02$ and 13 TeV.}
\begin{tabular}{ccccccccc}
\hline\hline
\multirow{2}{*}{Source of uncertainty} &   \multicolumn{8}{c}{Systematic uncertainty at $\eta=0$ (\%)} \\ 
&  \multicolumn{2}{c}{$\inelf$} &  \multicolumn{2}{c}{$\inels$} & \multicolumn{2}{c}{$\inelt$ }& \multicolumn{2}{c}{$\inelq$}    \\
\hline\hline
$\cms$ (TeV) & 5.02 & 13 & 5.02 & 13 & 5.02 & 13 & 5.02 & 13 \\
\hline
\hline
Track selection & $\pm2.0$ & $\pm2.0$ & $\pm2.5$ & $\pm2.3$ & $\pm1.8$& $\pm1.8$ & $\pm1.4$ & $\pm0.9$ \\
\hline
Event generator dependence & $\pm1$ & $\pm0.9$ & $\pm1.2$  & $\pm1.3$ & $\pm2$  & $\pm1.9$ & $\pm2$  & $\pm1.7$ \\
\hline
Acceptance and efficiency & $\pm1$  & $\pm0.5$ & $\pm1$  & $\pm0.6$ & $\pm1.3$  & $\pm0.7$ & $\pm1.8$  & $\pm1.8$\\
\hline
$z_{\mathrm{vtx}}$ dependence & $\pm0.3$ &  $\pm 0.2$ & $\pm0.3$   & $\pm 0.3$ & $\pm0.3$ & $\pm 0.3$ & $\pm0.5$  & $\pm 0.5$\\
\hline
Strangeness enhancement factor & $\pm0.3$  & $\pm 0.3$ & $\pm0.3$  &  $\pm 0.2$ & $\pm0.1$  & $\pm 0.1$ & neg. & neg. \\
\hline
Particle composition & $\pm 0.8$  & $\pm 0.7$ & $\pm 0.8$&  $\pm 0.8$ & $\pm 1$ & $\pm 1$ & $\pm 1.8$ & $\pm1.6$ \\
\hline
Material budget &$\pm 0.2$ & $\pm 0.2$ &$\pm 0.2$ & $\pm0.2$ & $\pm0.2$  & $\pm 0.2$ & $\pm 0.2$& $\pm0.2$\\
\hline\hline
Total systematic uncertainty & $\pm2.6$  & $\pm2.4$ & $\pm3.1$ & $\pm2.9$  &  $\pm3.2$ & $\pm2.9$ & $\pm3.6$  & $\pm3.1$ \\
\hline\hline
\end{tabular}

\label{tab:dndetasystracks}
\end{table}%

Several sources of systematic uncertainties were investigated. 
For the tracklet analysis at $\cms = 5.02$ TeV, the uncertainty related to the contribution of SD/DD events was evaluated by varying the fractions of SD and DD processes produced by PYTHIA 8 by $\pm 50\%$ of their nominal values.
The result of $\dndeta$ corrected by a track-to-particle correction map implementing the re-weighted SD mass distribution is used to determine the central values of the minimum-bias events. Results with two-times steeper re-weighted mass distribution and the default mass distribution in models were used to estimate the systematic uncertainty coming from the unknown SD mass distribution~\cite{Abelev:2012sea}. 
The highest deviation was taken as a systematic uncertainty. The third dominant source of uncertainty, which applies only to the tracklet analysis, includes the extrapolation of the number of particles as a function of $\pT$ from \unit[50]{MeV/$c$} down to zero, where the SPD is insensitive. The number of primary charged particles in this low $\pT$ range is varied conservatively in the event generator by $+100$\% and $-50$\%, adopted from the previous study~\cite{ALICE:2015olq}. The corresponding uncertainty is found to be $+1\%$ and $-0.5\%$ consistently for the three event classes.

For the analyses with a minimum $\pT$ threshold, the uncertainty due to the track selection is estimated by varying all the criteria around their nominal values. 
Three more sets of track selection criteria called tight- and loose-cut global tracks, and hybrid tracks~\cite{hybridExplanation} are considered in this study. The tight- and loose-cut global tracks are selected by tightening and loosening the DCA cuts with respect to the primary vertex, respectively. Hybrid tracks are composed of two types of track classes. The first class consists of tracks that have at least one hit in the SPD. The tracks from the second class do not have any SPD-associated hit and, hence, use the primary vertex as the innermost constraint for the tracking. The selection of hybrid tracks ensures a uniform distribution of tracks as a function of azimuthal angle.
This systematic uncertainty contribution increases from 2.8\% to 5\% with increasing $\pTcut$ and is slightly larger for the results at $\sqrt{s} = $ 5.02 TeV.

All other sources of uncertainty were estimated in the same way for tracklets and global tracks; all values can be found in Tables~\ref{tab:dndetasystracklets} and~\ref{tab:dndetasystracks}. 
The systematic uncertainty from the event generator dependence was also included. Data were corrected with PYTHIA 6~\cite{Sjostrand:2006za, Skands:2010ak}, and the relative deviation of the final result corrected with PYTHIA 8 was assigned as a systematic uncertainty. 
The uncertainty on the detector acceptance and efficiency was estimated by measuring $\mathrm{d}N_\mathrm{ch}/\mathrm{d}\eta$ for three different azimuthal regions and comparing it to the measurement in the whole region. The uncertainty due to the acceptance was studied additionally by dividing the whole event sample into ten different intervals of the primary vertex position within $z_\mathrm{vtx} = \pm10$ cm, each having the same number of events and performing the same analysis in each interval. 
The material budget in the ALICE central barrel is known to a precision of about 4$\%$~\cite{Abelev:2014ffa}. The corresponding systematic uncertainty, obtained by varying the material budget in the simulation, is estimated to be 0.2$\%$. 
The uncertainty associated with the correction for the difference in strange particle content between data and MC was estimated by varying the strange particle content in the simulation.
The yield of strange particles in data as compared to the simulation was measured as a function of $p_\mathrm{T}$ to be up to a factor 1.5 different, and this factor was varied by $\pm 30\%$.
Additionally, the particle composition affects the efficiency estimate because different particle species have different efficiencies that depend on the applied effective $p_\mathrm{T}$ cut-offs and the decay kinematics. The influence of this uncertainty was estimated by varying, in the simulation, the relative fraction of charged kaons, protons, and antiprotons with respect to charged pions by $\pm30\%$. 

The systematic uncertainties due to the event generator dependence, acceptance and efficiency, and $z_\mathrm{vtx}$ dependence are treated as uncorrelated between $\eta$ bins while the ones for the diffraction tuning (inclusive study only), zero-$\pT$ extrapolation (inclusive study only), strange particle abundance, particle composition, track selection ($\pT$ threshold study only), and material budget are correlated between $\eta$ bins in Fig.~\ref{fig:allmbresult}, ~\ref{fig:total_5TeV}, and~\ref{fig:total_13TeV}. The systematic uncertainties are usually correlated between different collision energies and among $\pT$ thresholds. 
Note that the statistical uncertainties in these analyses are negligible. 

\section {Results}

\begin{figure}[htbp]
	\centering
	\includegraphics[width=\linewidth]{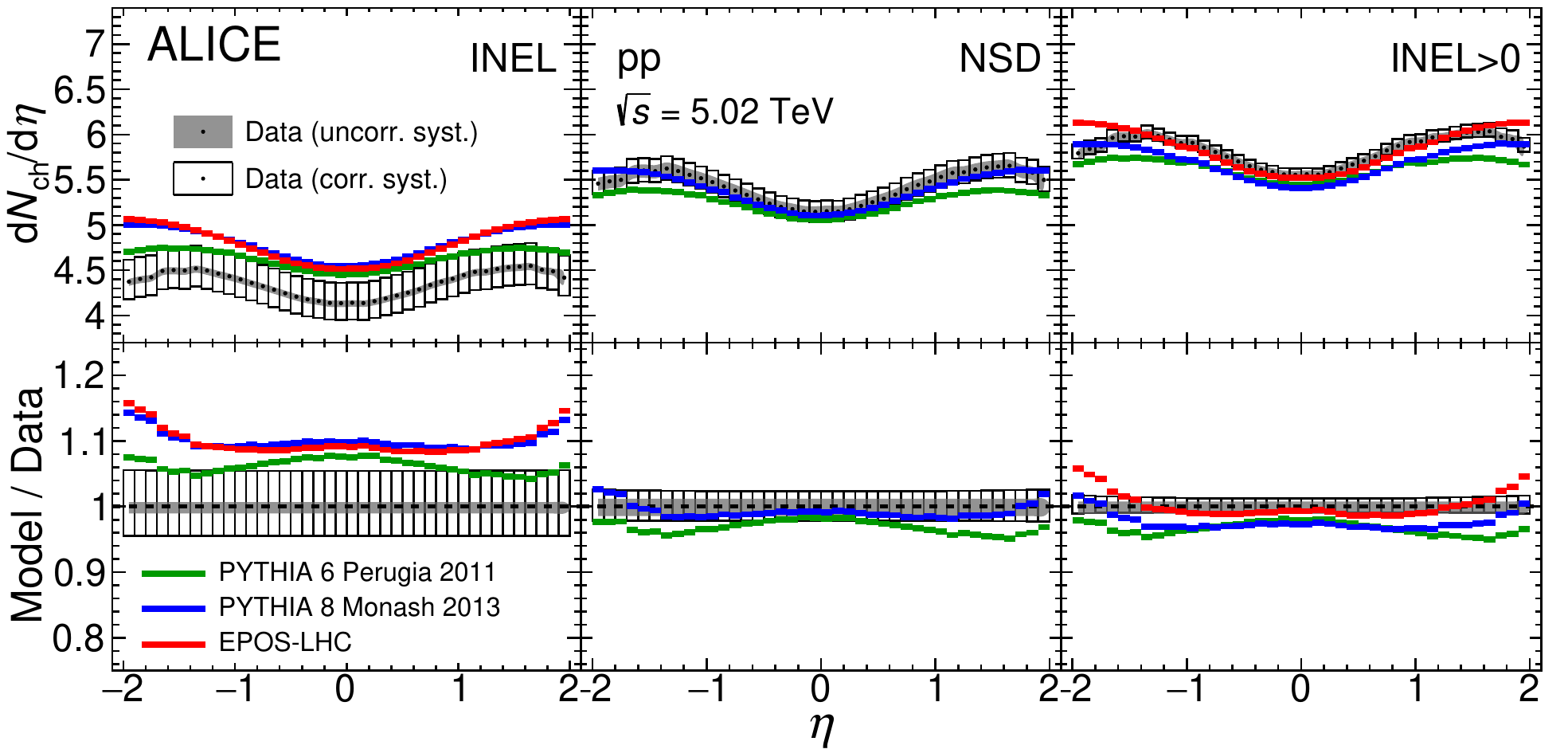}
	\caption{The distributions of $\dndeta$ for INEL (left panel), NSD (middle panel), and $\inelg$ (right panel) event classes in pp collisions at $\sqrt{s} = 5.02$ TeV. Data are compared to simulations obtained with PYTHIA 6 with the Perugia 2011 tuning, PYTHIA 8 with the Monash 2013 tuning, and EPOS-LHC. Grey bands (unfilled rectangles) represent the uncorrelated (correlated) systematic uncertainties from data. The bottom part of the figure shows the ratios between models and data.}
	\label{fig:allmbresult}
\end{figure}

The measurements of $\dndeta$ as a function of $\eta$ at $\sqrt{s} = 5.02$ TeV for INEL, NSD, and $\inelg$ events are shown in Fig.~\ref{fig:allmbresult}. The distributions of $\dndeta$ are compared to PYTHIA 6 with the Perugia 2011 tuning, PYTHIA 8 with the Monash 2013 tuning, and EPOS-LHC for the same event classes. 
In general, the models are better at describing the distributions which contain a smaller contribution from diffractive interactions. Therefore, the NSD and $\inelg$ event classes are described by models well. This can also be seen in the bottom panels of Fig.~\ref{fig:allmbresult} where the relative difference between models and data for the INEL event class stays within 10\% and for the NSD and $\inelg$ event classes stay within 5\%. 

The predictions of the two PYTHIA versions are very similar, however, PYTHIA 6 shows a better agreement with data for the INEL event class. EPOS-LHC describes the $\inelg$ events best at midrapidity, however, an increasing discrepancy emerges at forward rapidity. Note that EPOS-LHC is not provided for NSD events due to the models' lack of event type information. 
The values of the charged particle pseudorapidity density averaged over $|\eta|<0.5$ and $|\eta|<1$ ($\avdndeta$) are reported in Table~\ref{tab:averagedndeta}. The $\avdndeta$ values are provided for the INEL, NSD, and $\inelg$ event classes. The values obtained from the PYTHIA event generators are also reported.

\begin{table}[htbp]
\caption{The average $\dndeta$ ($\avdndeta$) in INEL, NSD, and $\inelg$ in pp collisions at $\cms = 5.02$~TeV.}
\centering
\renewcommand{\arraystretch}{1.2}
\begin{tabular}{c|cccccccc}
\hline\hline
\multirow{3}{*}{Event class} &   \multicolumn{8}{c}{ $\avdndeta$} \\ 
 &  \multicolumn{2}{c}{Data$\pm$syst.}  & \multicolumn{2}{c}{PYTHIA 6 Perugia} &  \multicolumn{2}{c}{PYTHIA 8 Monash} &  \multicolumn{2}{c}{EPOS-LHC}\\
 &  $|\eta|<0.5$ & $|\eta|<1$ & $|\eta|<0.5$ & $|\eta|<1$ & $|\eta|<0.5$ & $|\eta|<1$ & $|\eta|<0.5$ & $|\eta|<1$\\
\hline\hline
INEL & $4.17^{+0.23}_{-0.19}$ & $4.25^{+0.23}_{-0.19}$ & 4.48 & 4.54 & 4.58 & 4.65 & 4.54 & 4.62\\
\hline
NSD & $5.18^{+0.14}_{-0.13}$ & $5.28^{+0.13}_{-0.12}$ & 5.09 & 5.16 & 5.14 & 5.23 & N/A & N/A\\
\hline
$\inelg$ &  $5.60^{+0.08}_{-0.08}$ & $5.70^{+0.08}_{-0.07}$ &  5.48 & 5.55 & 5.44 & 5.54 & 5.55 & 5.65\\
\hline\hline
\end{tabular}

\label{tab:averagedndeta}
\end{table}

\begin{figure}[!htb]
\centering
\includegraphics[width=0.7\textwidth]{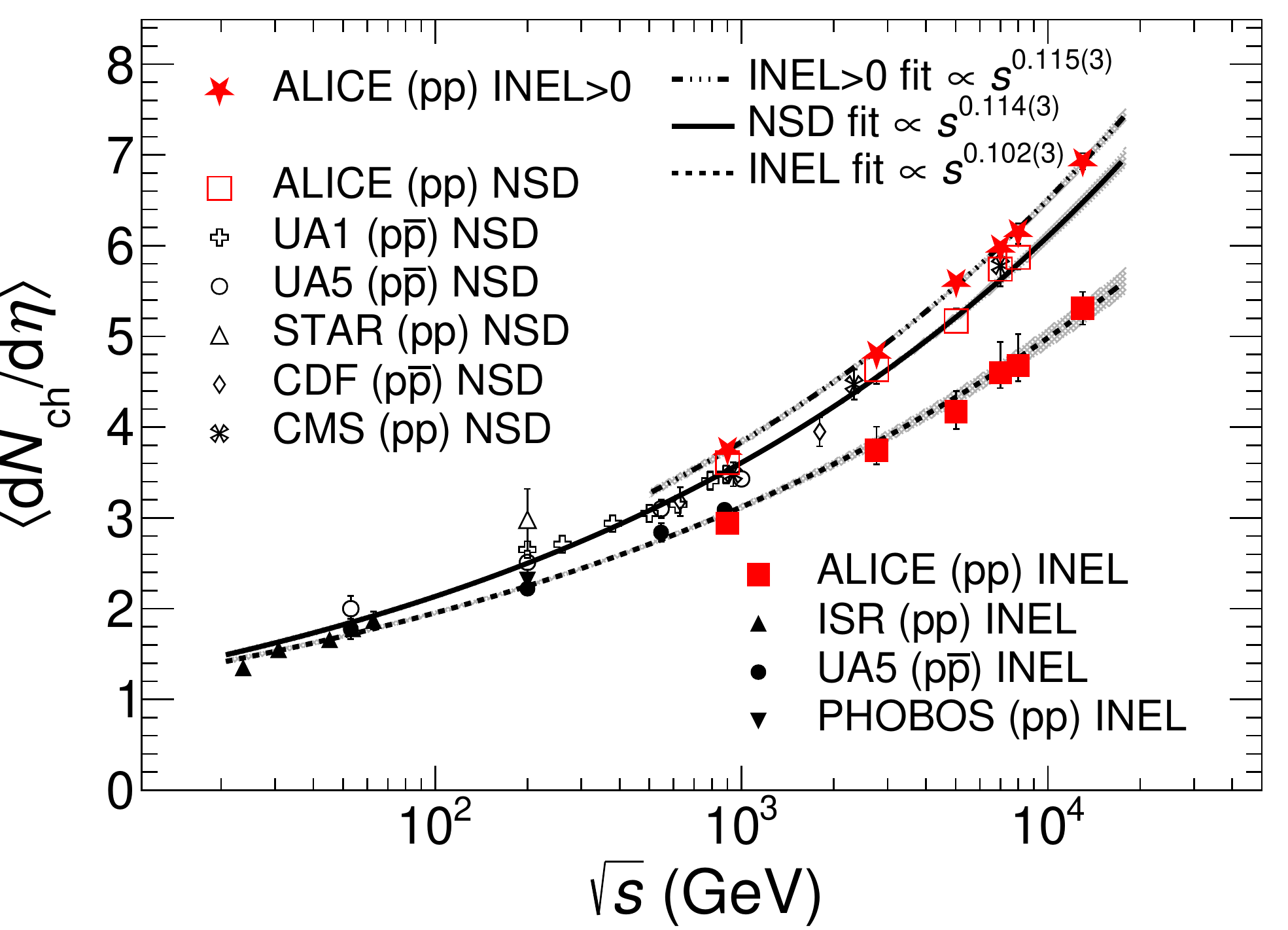}
\caption{The values of $\avdndeta$ averaged over $|\eta|<0.5$ for the INEL, NSD, and $\inelg$ event classes as a function of centre-of-mass energy~\cite{Aamodt:2009aa, Aamodt:2010ft, Aamodt:2010pp, ALICE:2015olq, Adam:2015pza, ALICE:2020swj,UA1:1989bou, UA5:1986yef, STAR:2008med, CDF:1989nkn, CMS:2010tjh, CMS:2010wcx, Ames-Bologna-CERN-Dortmund-Heidelberg-Warsaw:1983cqw, PHOBOS:2004xnp}. The lines indicate a power-law fit for each event class. The grey bands show one standard deviation of the fit.}
\label{fig:energydependence2}
\end{figure}

Figure~\ref{fig:energydependence2} shows the values of $\avdndeta$ averaged over $|\eta|<0.5$ for the INEL, NSD, and $\inelg$ event classes as a function of the centre-of-mass energy after combining the ALICE data~\cite{Aamodt:2009aa, Aamodt:2010ft, Aamodt:2010pp, ALICE:2015olq, Adam:2015pza, ALICE:2020swj} with other data at the LHC and at lower energies~\cite{UA1:1989bou, UA5:1986yef, STAR:2008med, CDF:1989nkn, CMS:2010tjh, CMS:2010wcx, Ames-Bologna-CERN-Dortmund-Heidelberg-Warsaw:1983cqw, PHOBOS:2004xnp}.  
At midrapidity, the measured $\dndeta$ can be parameterised by a power-law fit as $\dndeta \propto s^\delta$, resulting in $\delta = 0.102\pm0.003$, $0.114\pm0.003$, and $0.115\pm0.003$ for INEL, NSD, and $\inelg$ events, respectively. The energy dependence of particle production shows that the power-law fit is still valid. 

It is worth mentioning that the result for the INEL event class can be compared with the results in Pb--Pb and p--Pb collisions at the same centre-of-mass energies~\cite{ALICE:2012xs,ALICE:2015juo}.
These results can be compared to $\delta = 0.153 \pm 0.002 $ for central heavy-ion (A--A) collisions~\cite{Acharya:2018hhy,Basu:2020jbk}. This shows that the primary charged particle pseudorapidity density increases faster with energy in central A--A collisions compared to pp collisions, indicating that the initial longitudinal energy is more efficiently converted into particles in heavy-ion collisions relative to pp and p--Pb collisions. 

\begin{figure}[htbp]
	\centering
	\includegraphics[width=0.9\linewidth]{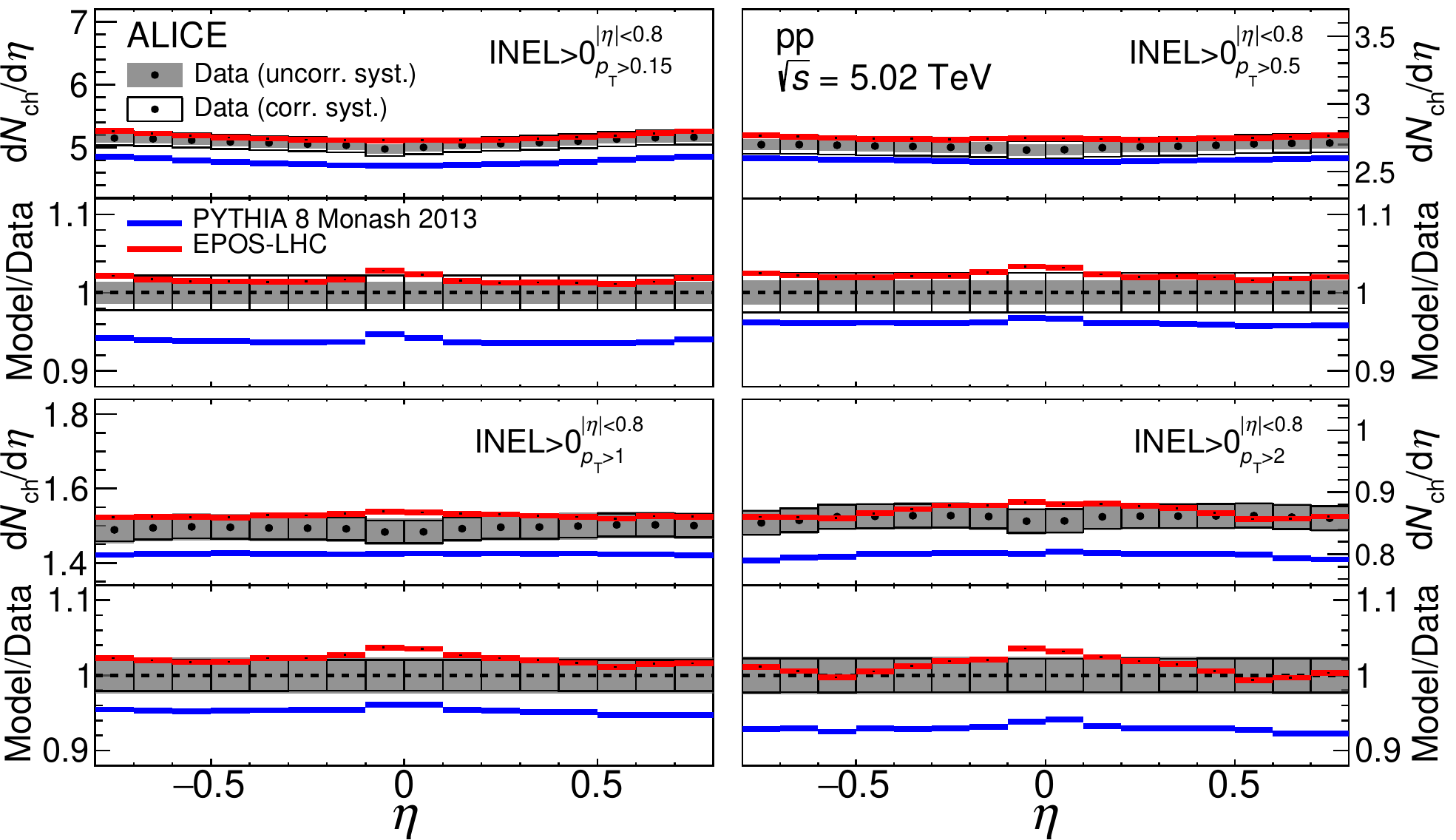}
	\caption{Pseudorapidity density distributions of charged particles, $\mathrm{d}N_{\mathrm{ch}}/\mathrm{d}\eta$, in pp collisions at $\sqrt{s} = $ 5.02 TeV for the four event classes, $\inelf$, $\inels$, $\inelt$, and $\inelq$, compared to the distributions from models: PYTHIA 8 Monash 2013 and EPOS-LHC. Grey bands (unfilled rectangles) represent the uncorrelated (correlated) systematic uncertainties from data.}
	\label{fig:total_5TeV}
\end{figure}

\begin{figure}[htbp]
	\centering
	\includegraphics[width=0.9\linewidth]{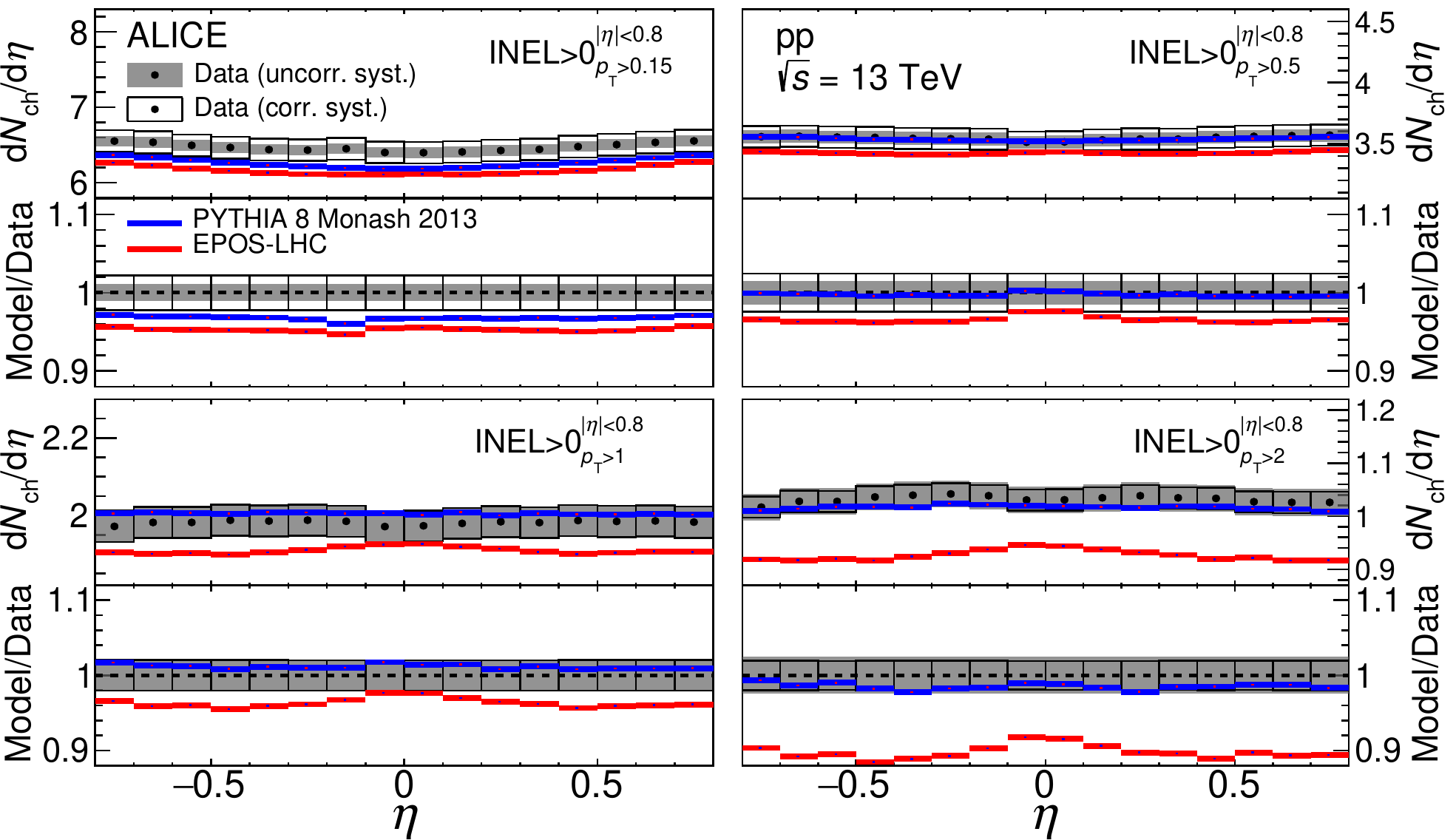}
	\caption{Pseudorapidity density distributions of charged particles, $\mathrm{d}N_{\mathrm{ch}}/\mathrm{d}\eta$, in pp collisions at $\sqrt{s}=$ 13 TeV for the four event classes, $\inelf$, $\inels$, $\inelt$, and $\inelq$, compared to the distributions from models: PYTHIA 8 Monash 2013 and EPOS-LHC. Grey bands (unfilled rectangles) represent the uncorrelated (correlated) systematic uncertainties from data.}
	\label{fig:total_13TeV}
\end{figure}

The measurements of $\dndeta$ as a function of $\eta$ at $\sqrt{s} = 5.02$ and 13 TeV are shown in Figs.~\ref{fig:total_5TeV} and~\ref{fig:total_13TeV}, for the $\inelf$, $\inels$, $\inelt$, and $\inelq$ event classes. The results are also compared to the predictions from the PYTHIA 8 with the Monash 2013 tuning and EPOS-LHC event generators, where EPOS-LHC was tuned on LHC Run 1 data at lower $\sqrt{s}$~\cite{Pierog:2013ria}. In general, the largest disagreement between data and MC is observed for the softest ($\inelf$) and hardest ($\inelq$) event classes. 
For collisions at $\sqrt{s} = 5.02$ TeV, EPOS-LHC describes the data to within 2\% in the measured $\eta$ range for all event classes, while PYTHIA 8 underestimates the measured $\dNdeta$ by 4--8\% depending on the $\pTcut$. On the other hand, at $\sqrt{s} =$ 13 TeV, PYTHIA 8 provides a better description of the measurements as compared to EPOS-LHC. At the highest collision energy, PYTHIA 8 predictions are consistent with the data within the uncertainties, while EPOS-LHC undershoots the data  by 4--10\% depending on the event class.
The results are expected to provide further constraints on charged particle production mechanisms implemented in models affecting both soft and hard QCD and their energy dependence.

\begin{figure}[htbp]
     \centering
	\includegraphics[width=\linewidth]{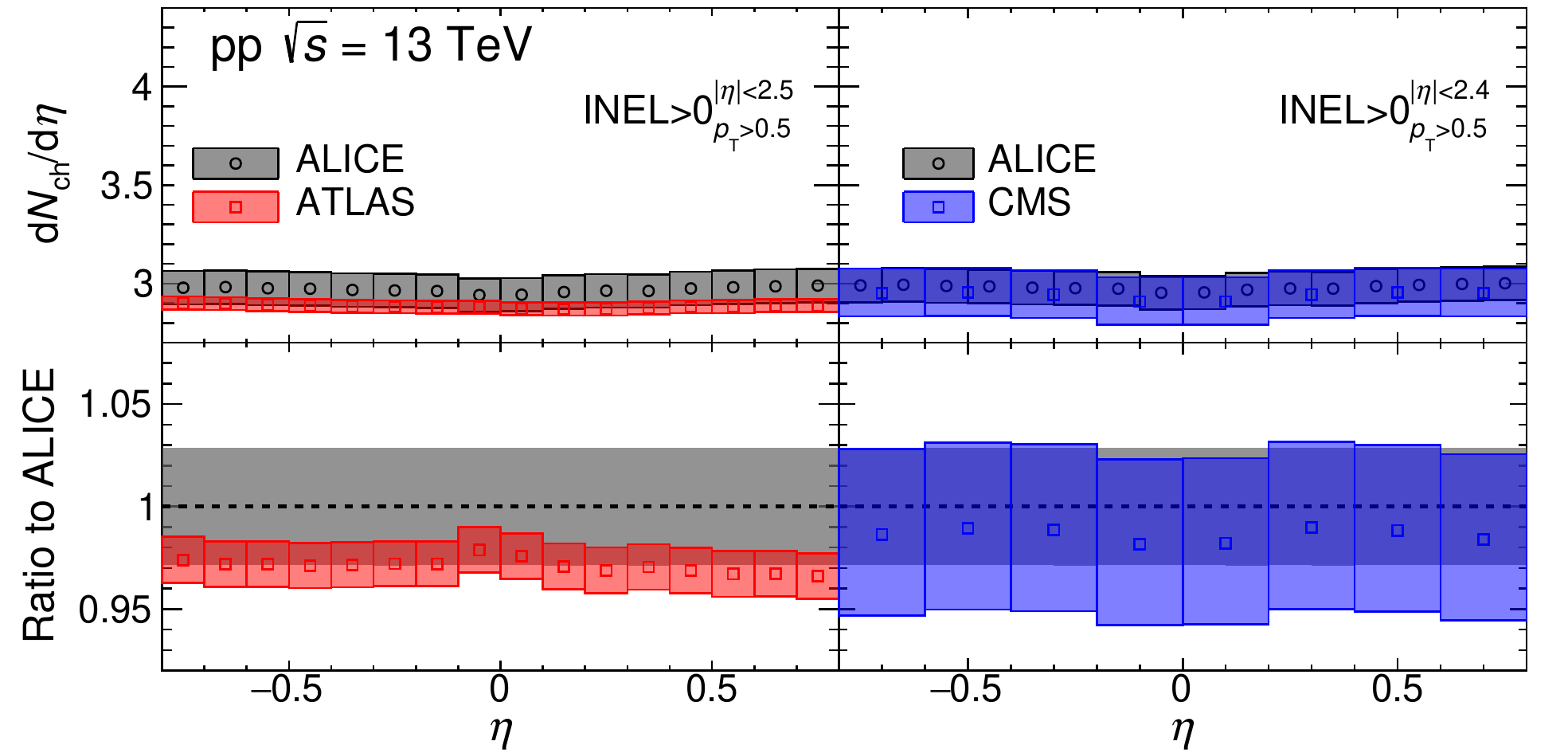}
        \caption{The distributions of $\dndeta$ for the $\inels$ event class are normalised to the $\inelsss$ (left) and $\inelss$ (right) event classes using PYTHIA 8 with the Monash 2013 tuning in pp collisions at $\cms = 13$ TeV~\cite{Aad:2016mok,CMS:2018nhd}. The bottom panels show the ratio of $\dndeta$ for the $\inelsss$ (left) and $\inelss$ (right) event class between ALICE and ATLAS (left) and between ALICE and CMS (right), respectively.}
        \label{fig:comparingtocmsatlas}
\end{figure}

The left panel of Fig.~\ref{fig:comparingtocmsatlas} shows the results for the $\inels$ event class extrapolated to the pseudorapidity interval $|\eta|<2.5$ to compare them to the ATLAS results~\cite{Aad:2016mok}. The normalisation factor is computed as the ratio of the $\dndeta$ for the $\inels$ and $\inelsss$ event classes obtained from PYTHIA 8 Monash 2013 simulations.
This normalisation is needed to do a correct comparison among different experiments because the condition with at least one charged particle depends on the acceptance. When experimentally the $\inelg$ condition is requested with a wider pseudorapidity acceptance, the corresponding event class is more inclusive because it collects more soft events. 
In the right panel of Fig.~\ref{fig:comparingtocmsatlas}, the same procedure is applied to normalise to the $\inelss$ event class in order to compare the result from ALICE to that obtained by CMS~\cite{CMS:2018nhd}. The result of ALICE is larger than those of ATLAS by $\sim3\%$ and CMS by up to $\sim2\%$. However, they are compatible within systematic uncertainties.

\section {Conclusions}

This article presents a set of measurements of the pseudorapidity density of primary charged particles ($\dndeta$) in proton--proton collisions at centre-of-mass energies $\cms = 5.02$ and 13 TeV. Results for inelastic (INEL) and non-single-diffractive (NSD) events as well as for inelastic events having at least one charged particle produced in the pseudorapidity interval $|\eta|<1$ ($\inelg$), are presented specifically at $\cms = 5.02$ TeV. The predictions of PYTHIA 6 with the Perugia 2011 tuning and PYTHIA 8 with the Monash 2013 tuning are close to each other. The two models show agreement with data except for the case of the NSD event class. Also, the result of the INEL event class is not well described by models due to the higher diffractive content.

The values of the average pseudorapidity density $\avdndeta$ in $|\eta|<0.5$ for INEL, NSD, and $\inelg$ events with an in-depth study for the single and double diffractive contributions are reported: $4.17^{+0.23}_{-0.19}$, $5.18^{+0.14}_{-0.13}$ and $5.60^{+0.08}_{-0.08}$ with systematic uncertainties, respectively.
The energy dependence of $\avdndeta$ is updated with the new values at $\sqrt{s} = 5.02$ TeV. Then, it is parameterised by a power-law fit as $\avdndeta \propto s^\delta$, resulting in $\delta = 0.102\pm 0.003$, $0.114\pm 0.003$, and $0.115\pm 0.004$ for INEL, NSD, and $\inelg$ events, respectively.

To provide detailed constraints on the charged particle production with hard processes in pp collisions at $\cms = 5.02$ and 13 TeV, the study is extended with the pseudorapidity density of primary charged particles in the pseudorapidity interval $|\eta|<0.8$ with minimum transverse momentum thresholds of \pT ~= 0.15, 0.5, 1, and 2 GeV/$c$ that are called $\inelf$, $\inels$, $\inelt$, and $\inelq$, respectively. The results of the $\dndeta$ distributions are also compared to the predictions from the PYTHIA 8 with the Monash 2013 tuning and EPOS-LHC event generators. PYTHIA 8 tends to underestimate the overall distributions at $\sqrt{s} = 5.02$ TeV by up to 8\%, while EPOS-LHC undershoots the measured multiplicities by up to 10\% as the $\pT$ threshold increases at $\sqrt{s}=13$ TeV. The largest disagreement between data and MC is observed for the softest ($\inelf$) and hardest ($\inelq$) event classes indicating the importance of these measurements to constrain models. 

In order to compare the ALICE result with minimum $\pT$ thresholds to those from the ATLAS and CMS experiments, the $\inels$ measurement is normalised to the $\inelsss$ and $\inelss$ event classes, respectively, using PYTHIA 8. The ALICE measurements agree with those from the other LHC experiments within systematic uncertainties.

\newenvironment{acknowledgement}{\relax}{\relax}
\begin{acknowledgement}
\section*{Acknowledgements}
\input{fa_2022-11-05_Opt_C.tex}
\end{acknowledgement}

\bibliographystyle{utphys}   
\bibliography{bibliography}

\newpage
\appendix

\section{The ALICE Collaboration}
\label{app:collab}
\input{2022-11-05-Alice_Authorlist_2022-11-05_Opt_C.tex}

\end{document}

%% file: fa_2022-11-05_Opt_C.tex

The ALICE Collaboration would like to thank all its engineers and technicians for their invaluable contributions to the construction of the experiment and the CERN accelerator teams for the outstanding performance of the LHC complex.
The ALICE Collaboration gratefully acknowledges the resources and support provided by all Grid centres and the Worldwide LHC Computing Grid (WLCG) collaboration.
The ALICE Collaboration acknowledges the following funding agencies for their support in building and running the ALICE detector:
A. I. Alikhanyan National Science Laboratory (Yerevan Physics Institute) Foundation (ANSL), State Committee of Science and World Federation of Scientists (WFS), Armenia;
Austrian Academy of Sciences, Austrian Science Fund (FWF): [M 2467-N36] and Nationalstiftung f\"{u}r Forschung, Technologie und Entwicklung, Austria;
Ministry of Communications and High Technologies, National Nuclear Research Center, Azerbaijan;
Conselho Nacional de Desenvolvimento Cient\'{\i}fico e Tecnol\'{o}gico (CNPq), Financiadora de Estudos e Projetos (Finep), Funda\c{c}\~{a}o de Amparo \`{a} Pesquisa do Estado de S\~{a}o Paulo (FAPESP) and Universidade Federal do Rio Grande do Sul (UFRGS), Brazil;
Bulgarian Ministry of Education and Science, within the National Roadmap for Research Infrastructures 2020-2027 (object CERN), Bulgaria;
Ministry of Education of China (MOEC) , Ministry of Science \& Technology of China (MSTC) and National Natural Science Foundation of China (NSFC), China;
Ministry of Science and Education and Croatian Science Foundation, Croatia;
Centro de Aplicaciones Tecnol\'{o}gicas y Desarrollo Nuclear (CEADEN), Cubaenerg\'{\i}a, Cuba;
Ministry of Education, Youth and Sports of the Czech Republic, Czech Republic;
The Danish Council for Independent Research | Natural Sciences, the VILLUM FONDEN and Danish National Research Foundation (DNRF), Denmark;
Helsinki Institute of Physics (HIP), Finland;
Commissariat \`{a} l'Energie Atomique (CEA) and Institut National de Physique Nucl\'{e}aire et de Physique des Particules (IN2P3) and Centre National de la Recherche Scientifique (CNRS), France;
Bundesministerium f\"{u}r Bildung und Forschung (BMBF) and GSI Helmholtzzentrum f\"{u}r Schwerionenforschung GmbH, Germany;
General Secretariat for Research and Technology, Ministry of Education, Research and Religions, Greece;
National Research, Development and Innovation Office, Hungary;
Department of Atomic Energy Government of India (DAE), Department of Science and Technology, Government of India (DST), University Grants Commission, Government of India (UGC) and Council of Scientific and Industrial Research (CSIR), India;
National Research and Innovation Agency - BRIN, Indonesia;
Istituto Nazionale di Fisica Nucleare (INFN), Italy;
Japanese Ministry of Education, Culture, Sports, Science and Technology (MEXT) and Japan Society for the Promotion of Science (JSPS) KAKENHI, Japan;
Consejo Nacional de Ciencia (CONACYT) y Tecnolog\'{i}a, through Fondo de Cooperaci\'{o}n Internacional en Ciencia y Tecnolog\'{i}a (FONCICYT) and Direcci\'{o}n General de Asuntos del Personal Academico (DGAPA), Mexico;
Nederlandse Organisatie voor Wetenschappelijk Onderzoek (NWO), Netherlands;
The Research Council of Norway, Norway;
Commission on Science and Technology for Sustainable Development in the South (COMSATS), Pakistan;
Pontificia Universidad Cat\'{o}lica del Per\'{u}, Peru;
Ministry of Education and Science, National Science Centre and WUT ID-UB, Poland;
Korea Institute of Science and Technology Information and National Research Foundation of Korea (NRF), Republic of Korea;
Ministry of Education and Scientific Research, Institute of Atomic Physics, Ministry of Research and Innovation and Institute of Atomic Physics and University Politehnica of Bucharest, Romania;
Ministry of Education, Science, Research and Sport of the Slovak Republic, Slovakia;
National Research Foundation of South Africa, South Africa;
Swedish Research Council (VR) and Knut \& Alice Wallenberg Foundation (KAW), Sweden;
European Organization for Nuclear Research, Switzerland;
Suranaree University of Technology (SUT), National Science and Technology Development Agency (NSTDA) and National Science, Research and Innovation Fund (NSRF via PMU-B B05F650021), Thailand;
Turkish Energy, Nuclear and Mineral Research Agency (TENMAK), Turkey;
National Academy of  Sciences of Ukraine, Ukraine;
Science and Technology Facilities Council (STFC), United Kingdom;
National Science Foundation of the United States of America (NSF) and United States Department of Energy, Office of Nuclear Physics (DOE NP), United States of America.
In addition, individual groups or members have received support from:
Marie Sk\l{}odowska Curie, European Research Council, Strong 2020 - Horizon 2020 (grant nos. 950692, 824093, 896850), European Union;
Academy of Finland (Center of Excellence in Quark Matter) (grant nos. 346327, 346328), Finland;
Programa de Apoyos para la Superaci\'{o}n del Personal Acad\'{e}mico, UNAM, Mexico.

%% file: 2022-11-05-Alice_Authorlist_2022-11-05_Opt_C.tex
\begin{flushleft} 
\small

S.~Acharya\,\orcidlink{0000-0002-9213-5329}\,$^{\rm 124}$, 
D.~Adamov\'{a}\,\orcidlink{0000-0002-0504-7428}\,$^{\rm 85}$, 
A.~Adler$^{\rm 69}$, 
G.~Aglieri Rinella\,\orcidlink{0000-0002-9611-3696}\,$^{\rm 32}$, 
M.~Agnello\,\orcidlink{0000-0002-0760-5075}\,$^{\rm 29}$, 
N.~Agrawal\,\orcidlink{0000-0003-0348-9836}\,$^{\rm 50}$, 
Z.~Ahammed\,\orcidlink{0000-0001-5241-7412}\,$^{\rm 132}$, 
S.~Ahmad$^{\rm 15}$, 
S.U.~Ahn\,\orcidlink{0000-0001-8847-489X}\,$^{\rm 70}$, 
I.~Ahuja\,\orcidlink{0000-0002-4417-1392}\,$^{\rm 37}$
% ...

\section*{Affiliation Notes}

$^{\rm I}$ Deceased\\
$^{\rm II}$ Also at: Max-Planck-Institut f\"{u}r Physik, Munich, Germany\\
$^{\rm III}$ Also at: Italian National Agency for New Technologies, Energy and Sustainable Economic Development (ENEA), Bologna, Italy\\
$^{\rm IV}$ Also at: Dipartimento DET del Politecnico di Torino, Turin, Italy\\
$^{\rm V}$ Also at: Department of Applied Physics, Aligarh Muslim University, Aligarh, India\\
$^{\rm VI}$ Also at: Institute of Theoretical Physics, University of Wroclaw, Poland\\
$^{\rm VII}$ Also at: An institution covered by a cooperation agreement with CERN\\

\section*{Collaboration Institutes}

$^{1}$ A.I. Alikhanyan National Science Laboratory (Yerevan Physics Institute) Foundation, Yerevan, Armenia\\
$^{2}$ AGH University of Krakow, Cracow, Poland\\
$^{3}$ Bogolyubov Institute for Theoretical Physics, National Academy of Sciences of Ukraine, Kiev, Ukraine\\
$^{4}$ Bose Institute, Department of Physics  and Centre for Astroparticle Physics and Space Science (CAPSS), Kolkata, India\\
$^{5}$ California Polytechnic State University, San Luis Obispo, California, United States\\
$^{6}$ Central China Normal University, Wuhan, China\\
$^{7}$ Centro de Aplicaciones Tecnol\'{o}gicas y Desarrollo Nuclear (CEADEN), Havana, Cuba\\
$^{8}$ Centro de Investigaci\'{o}n y de Estudios Avanzados (CINVESTAV), Mexico City and M\'{e}rida, Mexico\\
$^{9}$ Chicago State University, Chicago, Illinois, United States\\
$^{10}$ China Institute of Atomic Energy, Beijing, China\\
$^{11}$ Chungbuk National University, Cheongju, Republic of Korea\\
$^{12}$ Comenius University Bratislava, Faculty of Mathematics, Physics and Informatics, Bratislava, Slovak Republic\\
$^{13}$ COMSATS University Islamabad, Islamabad, Pakistan\\
$^{14}$ Creighton University, Omaha, Nebraska, United States\\
$^{15}$ Department of Physics, Aligarh Muslim University, Aligarh, India\\
$^{16}$ Department of Physics, Pusan National University, Pusan, Republic of Korea\\
$^{17}$ Department of Physics, Sejong University, Seoul, Republic of Korea\\
$^{18}$ Department of Physics, University of California, Berkeley, California, United States\\
$^{19}$ Department of Physics, University of Oslo, Oslo, Norway\\
$^{20}$ Department of Physics and Technology, University of Bergen, Bergen, Norway\\
$^{21}$ Dipartimento di Fisica, Universit\`{a} di Pavia, Pavia, Italy\\
$^{22}$ Dipartimento di Fisica dell'Universit\`{a} and Sezione INFN, Cagliari, Italy\\
$^{23}$ Dipartimento di Fisica dell'Universit\`{a} and Sezione INFN, Trieste, Italy\\
$^{24}$ Dipartimento di Fisica dell'Universit\`{a} and Sezione INFN, Turin, Italy\\
$^{25}$ Dipartimento di Fisica e Astronomia dell'Universit\`{a} and Sezione INFN, Bologna, Italy\\
$^{26}$ Dipartimento di Fisica e Astronomia dell'Universit\`{a} and Sezione INFN, Catania, Italy\\
$^{27}$ Dipartimento di Fisica e Astronomia dell'Universit\`{a} and Sezione INFN, Padova, Italy\\
$^{28}$ Dipartimento di Fisica `E.R.~Caianiello' dell'Universit\`{a} and Gruppo Collegato INFN, Salerno, Italy\\
$^{29}$ Dipartimento DISAT del Politecnico and Sezione INFN, Turin, Italy\\
$^{30}$ Dipartimento di Scienze MIFT, Universit\`{a} di Messina, Messina, Italy\\
$^{31}$ Dipartimento Interateneo di Fisica `M.~Merlin' and Sezione INFN, Bari, Italy\\
$^{32}$ European Organization for Nuclear Research (CERN), Geneva, Switzerland\\
$^{33}$ Faculty of Electrical Engineering, Mechanical Engineering and Naval Architecture, University of Split, Split, Croatia\\
$^{34}$ Faculty of Engineering and Science, Western Norway University of Applied Sciences, Bergen, Norway\\
$^{35}$ Faculty of Nuclear Sciences and Physical Engineering, Czech Technical University in Prague, Prague, Czech Republic\\
$^{36}$ Faculty of Physics, Sofia University, Sofia, Bulgaria\\
$^{37}$ Faculty of Science, P.J.~\v{S}af\'{a}rik University, Ko\v{s}ice, Slovak Republic\\
$^{38}$ Frankfurt Institute for Advanced Studies, Johann Wolfgang Goethe-Universit\"{a}t Frankfurt, Frankfurt, Germany\\
$^{39}$ Fudan University, Shanghai, China\\
$^{40}$ Gangneung-Wonju National University, Gangneung, Republic of Korea\\
$^{41}$ Gauhati University, Department of Physics, Guwahati, India\\
$^{42}$ Helmholtz-Institut f\"{u}r Strahlen- und Kernphysik, Rheinische Friedrich-Wilhelms-Universit\"{a}t Bonn, Bonn, Germany\\
$^{43}$ Helsinki Institute of Physics (HIP), Helsinki, Finland\\
$^{44}$ High Energy Physics Group,  Universidad Aut\'{o}noma de Puebla, Puebla, Mexico\\
$^{45}$ Horia Hulubei National Institute of Physics and Nuclear Engineering, Bucharest, Romania\\
$^{46}$ Indian Institute of Technology Bombay (IIT), Mumbai, India\\
$^{47}$ Indian Institute of Technology Indore, Indore, India\\
$^{48}$ INFN, Laboratori Nazionali di Frascati, Frascati, Italy\\
$^{49}$ INFN, Sezione di Bari, Bari, Italy\\
$^{50}$ INFN, Sezione di Bologna, Bologna, Italy\\
$^{51}$ INFN, Sezione di Cagliari, Cagliari, Italy\\
$^{52}$ INFN, Sezione di Catania, Catania, Italy\\
$^{53}$ INFN, Sezione di Padova, Padova, Italy\\
$^{54}$ INFN, Sezione di Pavia, Pavia, Italy\\
$^{55}$ INFN, Sezione di Torino, Turin, Italy\\
$^{56}$ INFN, Sezione di Trieste, Trieste, Italy\\
$^{57}$ Inha University, Incheon, Republic of Korea\\
$^{58}$ Institute for Gravitational and Subatomic Physics (GRASP), Utrecht University/Nikhef, Utrecht, Netherlands\\
$^{59}$ Institute of Experimental Physics, Slovak Academy of Sciences, Ko\v{s}ice, Slovak Republic\\
$^{60}$ Institute of Physics, Homi Bhabha National Institute, Bhubaneswar, India\\
$^{61}$ Institute of Physics of the Czech Academy of Sciences, Prague, Czech Republic\\
$^{62}$ Institute of Space Science (ISS), Bucharest, Romania\\
$^{63}$ Institut f\"{u}r Kernphysik, Johann Wolfgang Goethe-Universit\"{a}t Frankfurt, Frankfurt, Germany\\
$^{64}$ Instituto de Ciencias Nucleares, Universidad Nacional Aut\'{o}noma de M\'{e}xico, Mexico City, Mexico\\
$^{65}$ Instituto de F\'{i}sica, Universidade Federal do Rio Grande do Sul (UFRGS), Porto Alegre, Brazil\\
$^{66}$ Instituto de F\'{\i}sica, Universidad Nacional Aut\'{o}noma de M\'{e}xico, Mexico City, Mexico\\
$^{67}$ iThemba LABS, National Research Foundation, Somerset West, South Africa\\
$^{68}$ Jeonbuk National University, Jeonju, Republic of Korea\\
$^{69}$ Johann-Wolfgang-Goethe Universit\"{a}t Frankfurt Institut f\"{u}r Informatik, Fachbereich Informatik und Mathematik, Frankfurt, Germany\\
$^{70}$ Korea Institute of Science and Technology Information, Daejeon, Republic of Korea\\
$^{71}$ KTO Karatay University, Konya, Turkey\\
$^{72}$ Laboratoire de Physique Subatomique et de Cosmologie, Universit\'{e} Grenoble-Alpes, CNRS-IN2P3, Grenoble, France\\
$^{73}$ Lawrence Berkeley National Laboratory, Berkeley, California, United States\\
$^{74}$ Lund University Department of Physics, Division of Particle Physics, Lund, Sweden\\
$^{75}$ Nagasaki Institute of Applied Science, Nagasaki, Japan\\
$^{76}$ Nara Women{'}s University (NWU), Nara, Japan\\
$^{77}$ National and Kapodistrian University of Athens, School of Science, Department of Physics , Athens, Greece\\
$^{78}$ National Centre for Nuclear Research, Warsaw, Poland\\
$^{79}$ National Institute of Science Education and Research, Homi Bhabha National Institute, Jatni, India\\
$^{80}$ National Nuclear Research Center, Baku, Azerbaijan\\
$^{81}$ National Research and Innovation Agency - BRIN, Jakarta, Indonesia\\
$^{82}$ Niels Bohr Institute, University of Copenhagen, Copenhagen, Denmark\\
$^{83}$ Nikhef, National institute for subatomic physics, Amsterdam, Netherlands\\
$^{84}$ Nuclear Physics Group, STFC Daresbury Laboratory, Daresbury, United Kingdom\\
$^{85}$ Nuclear Physics Institute of the Czech Academy of Sciences, Husinec-\v{R}e\v{z}, Czech Republic\\
$^{86}$ Oak Ridge National Laboratory, Oak Ridge, Tennessee, United States\\
$^{87}$ Ohio State University, Columbus, Ohio, United States\\
$^{88}$ Physics department, Faculty of science, University of Zagreb, Zagreb, Croatia\\
$^{89}$ Physics Department, Panjab University, Chandigarh, India\\
$^{90}$ Physics Department, University of Jammu, Jammu, India\\
$^{91}$ Physics Program and International Institute for Sustainability with Knotted Chiral Meta Matter (SKCM2), Hiroshima University, Hiroshima, Japan\\
$^{92}$ Physikalisches Institut, Eberhard-Karls-Universit\"{a}t T\"{u}bingen, T\"{u}bingen, Germany\\
$^{93}$ Physikalisches Institut, Ruprecht-Karls-Universit\"{a}t Heidelberg, Heidelberg, Germany\\
$^{94}$ Physik Department, Technische Universit\"{a}t M\"{u}nchen, Munich, Germany\\
$^{95}$ Politecnico di Bari and Sezione INFN, Bari, Italy\\
$^{96}$ Research Division and ExtreMe Matter Institute EMMI, GSI Helmholtzzentrum f\"ur Schwerionenforschung GmbH, Darmstadt, Germany\\
$^{97}$ Saga University, Saga, Japan\\
$^{98}$ Saha Institute of Nuclear Physics, Homi Bhabha National Institute, Kolkata, India\\
$^{99}$ School of Physics and Astronomy, University of Birmingham, Birmingham, United Kingdom\\
$^{100}$ Secci\'{o}n F\'{\i}sica, Departamento de Ciencias, Pontificia Universidad Cat\'{o}lica del Per\'{u}, Lima, Peru\\
$^{101}$ Stefan Meyer Institut f\"{u}r Subatomare Physik (SMI), Vienna, Austria\\
$^{102}$ SUBATECH, IMT Atlantique, Nantes Universit\'{e}, CNRS-IN2P3, Nantes, France\\
$^{103}$ Sungkyunkwan University, Suwon City, Republic of Korea\\
$^{104}$ Suranaree University of Technology, Nakhon Ratchasima, Thailand\\
$^{105}$ Technical University of Ko\v{s}ice, Ko\v{s}ice, Slovak Republic\\
$^{106}$ The Henryk Niewodniczanski Institute of Nuclear Physics, Polish Academy of Sciences, Cracow, Poland\\
$^{107}$ The University of Texas at Austin, Austin, Texas, United States\\
$^{108}$ Universidad Aut\'{o}noma de Sinaloa, Culiac\'{a}n, Mexico\\
$^{109}$ Universidade de S\~{a}o Paulo (USP), S\~{a}o Paulo, Brazil\\
$^{110}$ Universidade Estadual de Campinas (UNICAMP), Campinas, Brazil\\
$^{111}$ Universidade Federal do ABC, Santo Andre, Brazil\\
$^{112}$ University of Cape Town, Cape Town, South Africa\\
$^{113}$ University of Houston, Houston, Texas, United States\\
$^{114}$ University of Jyv\"{a}skyl\"{a}, Jyv\"{a}skyl\"{a}, Finland\\
$^{115}$ University of Kansas, Lawrence, Kansas, United States\\
$^{116}$ University of Liverpool, Liverpool, United Kingdom\\
$^{117}$ University of Science and Technology of China, Hefei, China\\
$^{118}$ University of South-Eastern Norway, Kongsberg, Norway\\
$^{119}$ University of Tennessee, Knoxville, Tennessee, United States\\
$^{120}$ University of the Witwatersrand, Johannesburg, South Africa\\
$^{121}$ University of Tokyo, Tokyo, Japan\\
$^{122}$ University of Tsukuba, Tsukuba, Japan\\
$^{123}$ University Politehnica of Bucharest, Bucharest, Romania\\
$^{124}$ Universit\'{e} Clermont Auvergne, CNRS/IN2P3, LPC, Clermont-Ferrand, France\\
$^{125}$ Universit\'{e} de Lyon, CNRS/IN2P3, Institut de Physique des 2 Infinis de Lyon, Lyon, France\\
$^{126}$ Universit\'{e} de Strasbourg, CNRS, IPHC UMR 7178, F-67000 Strasbourg, France, Strasbourg, France\\
$^{127}$ Universit\'{e} Paris-Saclay, Centre d'Etudes de Saclay (CEA), IRFU, D\'{e}partment de Physique Nucl\'{e}aire (DPhN), Saclay, France\\
$^{128}$ Universit\'{e}  Paris-Saclay, CNRS/IN2P3, IJCLab, Orsay, France\\
$^{129}$ Universit\`{a} degli Studi di Foggia, Foggia, Italy\\
$^{130}$ Universit\`{a} del Piemonte Orientale, Vercelli, Italy\\
$^{131}$ Universit\`{a} di Brescia, Brescia, Italy\\
$^{132}$ Variable Energy Cyclotron Centre, Homi Bhabha National Institute, Kolkata, India\\
$^{133}$ Warsaw University of Technology, Warsaw, Poland\\
$^{134}$ Wayne State University, Detroit, Michigan, United States\\
$^{135}$ Westf\"{a}lische Wilhelms-Universit\"{a}t M\"{u}nster, Institut f\"{u}r Kernphysik, M\"{u}nster, Germany\\
$^{136}$ Wigner Research Centre for Physics, Budapest, Hungary\\
$^{137}$ Yale University, New Haven, Connecticut, United States\\
$^{138}$ Yonsei University, Seoul, Republic of Korea\\
$^{139}$  Zentrum  f\"{u}r Technologie und Transfer (ZTT), Worms, Germany\\
$^{140}$ Affiliated with an institute covered by a cooperation agreement with CERN\\
$^{141}$ Affiliated with an international laboratory covered by a cooperation agreement with CERN.\\

\end{flushleft}

%% file: main.bbl
\providecommand{\href}[2]{#2}\begingroup\raggedright\begin{thebibliography}{10}

\bibitem{Field:1976ve}
R.~D. Field and R.~P. Feynman, ``{Quark Elastic Scattering as a Source of High
  Transverse Momentum Mesons}'',
  \href{http://dx.doi.org/10.1103/PhysRevD.15.2590}{{\em Phys. Rev. D}
  {\bfseries 15} (1977) 2590--2616}.

\bibitem{Srednicki:2007qs}
M.~Srednicki, {\em {Quantum field theory}}.
\newblock Cambridge University Press, 1, 2007.

\bibitem{Collins:1977jy}
P.~D.~B. Collins, \href{http://dx.doi.org/10.1017/CBO9780511897603}{{\em {An
  Introduction to Regge Theory and High-Energy Physics}}}.
\newblock Cambridge Monographs on Mathematical Physics. Cambridge Univ. Press,
  Cambridge, UK, 5, 2009.

\bibitem{Greiner:2002ui}
W.~Greiner, S.~Schramm, and E.~Stein,
  \href{http://dx.doi.org/https://doi.org/10.1007/978-3-540-48535-3}{{\em
  {Quantum chromodynamics}}}.
\newblock Springer-Verlag Berlin, Heidelberg, 2002.

\bibitem{Ostapchenko:2007qb}
S.~Ostapchenko, ``{Status of QGSJET}'',
  \href{http://dx.doi.org/10.1063/1.2775904}{{\em AIP Conf. Proc.} {\bfseries
  928} (2007) 118--125}, \href{http://arxiv.org/abs/0706.3784}{{\ttfamily
  arXiv:0706.3784 [hep-ph]}}.

\bibitem{Fletcher:1994bd}
R.~S. Fletcher, T.~K. Gaisser, P.~Lipari, and T.~Stanev, ``{SIBYLL: An Event
  generator for simulation of high-energy cosmic ray cascades}'',
  \href{http://dx.doi.org/10.1103/PhysRevD.50.5710}{{\em Phys. Rev. D}
  {\bfseries 50} (1994) 5710--5731}.

\bibitem{Engel:1994vs}
R.~Engel, ``{Photoproduction within the two component dual parton model. 1.
  Amplitudes and cross-sections}'',
  \href{http://dx.doi.org/10.1007/BF01496594}{{\em Z. Phys. C} {\bfseries 66}
  (1995) 203--214}.

\bibitem{Sjostrand:2006za}
T.~Sjostrand, S.~Mrenna, and P.~Z. Skands, ``{PYTHIA 6.4 Physics and Manual}'',
  \href{http://dx.doi.org/10.1088/1126-6708/2006/05/026}{{\em JHEP} {\bfseries
  05} (2006) 026},
\href{http://arxiv.org/abs/hep-ph/0603175}{{\ttfamily arXiv:hep-ph/0603175
  [hep-ph]}}.

\bibitem{Pierog:2013ria}
T.~Pierog, I.~Karpenko, J.~M. Katzy, E.~Yatsenko, and K.~Werner, ``{EPOS LHC:
  Test of collective hadronization with data measured at the CERN Large Hadron
  Collider}'', \href{http://dx.doi.org/10.1103/PhysRevC.92.034906}{{\em Phys.
  Rev. C} {\bfseries 92} (2015) 034906},
  \href{http://arxiv.org/abs/1306.0121}{{\ttfamily arXiv:1306.0121 [hep-ph]}}.

\bibitem{Bahr:2008pv}
M.~Bahr {\em et~al.}, ``{Herwig++ Physics and Manual}'',
  \href{http://dx.doi.org/10.1140/epjc/s10052-008-0798-9}{{\em Eur. Phys. J. C}
  {\bfseries 58} (2008) 639--707},
  \href{http://arxiv.org/abs/0803.0883}{{\ttfamily arXiv:0803.0883 [hep-ph]}}.

\bibitem{Gleisberg:2008ta}
T.~Gleisberg, S.~Hoeche, F.~Krauss, M.~Schonherr, S.~Schumann, F.~Siegert, and
  J.~Winter, ``{Event generation with SHERPA 1.1}'',
  \href{http://dx.doi.org/10.1088/1126-6708/2009/02/007}{{\em JHEP} {\bfseries
  02} (2009) 007}, \href{http://arxiv.org/abs/0811.4622}{{\ttfamily
  arXiv:0811.4622 [hep-ph]}}.

\bibitem{Skands:2010ak}
P.~Z. Skands, ``{Tuning Monte Carlo Generators: The Perugia Tunes}'',
  \href{http://dx.doi.org/10.1103/PhysRevD.82.074018}{{\em Phys. Rev. D}
  {\bfseries 82} (2010) 074018},
  \href{http://arxiv.org/abs/1005.3457}{{\ttfamily arXiv:1005.3457 [hep-ph]}}.

\bibitem{Sjostrand:2007gs}
T.~Sjostrand, S.~Mrenna, and P.~Z. Skands, ``{A Brief Introduction to PYTHIA
  8.1}'', \href{http://dx.doi.org/10.1016/j.cpc.2008.01.036}{{\em
  Comput.Phys.Commun.} {\bfseries 178} (2008) 852--867},
\href{http://arxiv.org/abs/0710.3820}{{\ttfamily arXiv:0710.3820 [hep-ph]}}.

\bibitem{Sjostrand:2014zea}
T.~Sj\"ostrand {\em et~al.}, ``{An introduction to PYTHIA 8.2}'',
  \href{http://dx.doi.org/10.1016/j.cpc.2015.01.024}{{\em Comput. Phys.
  Commun.} {\bfseries 191} (2015) 159--177},
  \href{http://arxiv.org/abs/1410.3012}{{\ttfamily arXiv:1410.3012 [hep-ph]}}.

\bibitem{Skands:2014pea}
P.~Skands, S.~Carrazza, and J.~Rojo, ``{Tuning PYTHIA 8.1: the Monash 2013
  Tune}'', \href{http://dx.doi.org/10.1140/epjc/s10052-014-3024-y}{{\em Eur.
  Phys. J.} {\bfseries C74} (2014) 3024},
\href{http://arxiv.org/abs/1404.5630}{{\ttfamily arXiv:1404.5630 [hep-ph]}}.

\bibitem{Aamodt:2009aa}
{\bfseries ALICE} Collaboration, K.~Aamodt {\em et~al.}, ``{First proton-proton
  collisions at the LHC as observed with the ALICE detector: Measurement of the
  charged particle pseudorapidity density at $\sqrt{s}$ = 900 GeV}'',
  \href{http://dx.doi.org/10.1140/epjc/s10052-009-1227-4}{{\em Eur. Phys. J. C}
  {\bfseries 65} (2010) 111--125},
  \href{http://arxiv.org/abs/0911.5430}{{\ttfamily arXiv:0911.5430 [hep-ex]}}.

\bibitem{Aamodt:2010ft}
{\bfseries ALICE} Collaboration, K.~Aamodt {\em et~al.}, ``{Charged-particle
  multiplicity measurement in proton-proton collisions at $\sqrt{s}=0.9$ and
  2.36 TeV with ALICE at LHC}'',
  \href{http://dx.doi.org/10.1140/epjc/s10052-010-1339-x}{{\em Eur. Phys. J. C}
  {\bfseries 68} (2010) 89--108},
  \href{http://arxiv.org/abs/1004.3034}{{\ttfamily arXiv:1004.3034 [hep-ex]}}.

\bibitem{Aamodt:2010pp}
{\bfseries ALICE} Collaboration, K.~Aamodt {\em et~al.}, ``{Charged-particle
  multiplicity measurement in proton-proton collisions at $\sqrt{s}=7$ TeV with
  ALICE at LHC}'', \href{http://dx.doi.org/10.1140/epjc/s10052-010-1350-2}{{\em
  Eur. Phys. J. C} {\bfseries 68} (2010) 345--354},
  \href{http://arxiv.org/abs/1004.3514}{{\ttfamily arXiv:1004.3514 [hep-ex]}}.

\bibitem{ALICE:2015olq}
{\bfseries ALICE} Collaboration, J.~Adam {\em et~al.}, ``{Charged-particle
  multiplicities in proton\textendash{}proton collisions at $\sqrt{s} = 0.9$ to
  8 TeV}'', \href{http://dx.doi.org/10.1140/epjc/s10052-016-4571-1}{{\em Eur.
  Phys. J. C} {\bfseries 77} (2017) 33},
  \href{http://arxiv.org/abs/1509.07541}{{\ttfamily arXiv:1509.07541
  [nucl-ex]}}.

\bibitem{Adam:2015pza}
{\bfseries ALICE} Collaboration, J.~Adam {\em et~al.}, ``{Pseudorapidity and
  transverse-momentum distributions of charged particles in
  proton\textendash{}proton collisions at $\sqrt s=$ 13 TeV}'',
  \href{http://dx.doi.org/10.1016/j.physletb.2015.12.030}{{\em Phys. Lett. B}
  {\bfseries 753} (2016) 319--329},
  \href{http://arxiv.org/abs/1509.08734}{{\ttfamily arXiv:1509.08734
  [nucl-ex]}}.

\bibitem{ALICE:2020swj}
{\bfseries ALICE} Collaboration, S.~Acharya {\em et~al.}, ``{Pseudorapidity
  distributions of charged particles as a function of mid- and forward rapidity
  multiplicities in pp collisions at $\sqrt{s}$~=~5.02, 7 and 13 TeV}'',
  \href{http://dx.doi.org/10.1140/epjc/s10052-021-09349-5}{{\em Eur. Phys. J.
  C} {\bfseries 81} (2021) 630},
  \href{http://arxiv.org/abs/2009.09434}{{\ttfamily arXiv:2009.09434
  [nucl-ex]}}.

\bibitem{ALICE-PUBLIC-2017-005}
{\bfseries ALICE} Collaboration, ``{The ALICE definition of primary
  particles}'', {\em ALICE-PUBLIC-NOTE-2017-005} (2017) .
  \url{https://cds.cern.ch/record/2270008}.

\bibitem{ALICE:2013bva}
{\bfseries ALICE} Collaboration, ``{Charged-particle multiplicity measurement
  with Reconstructed Tracks in pp Collisions at $\sqrt{s}$ = 0.9 and 7 TeV with
  ALICE at the LHC}'', {\em ALICE-PUBLIC-2013-001} (7, 2013) .

\bibitem{Aad:2010ac}
{\bfseries ATLAS} Collaboration, G.~Aad {\em et~al.}, ``{Charged-particle
  multiplicities in pp interactions measured with the ATLAS detector at the
  LHC}'', \href{http://dx.doi.org/10.1088/1367-2630/13/5/053033}{{\em New J.
  Phys.} {\bfseries 13} (2011) 053033},
  \href{http://arxiv.org/abs/1012.5104}{{\ttfamily arXiv:1012.5104 [hep-ex]}}.

\bibitem{Aad:2016mok}
{\bfseries ATLAS} Collaboration, G.~Aad {\em et~al.}, ``{Charged-particle
  distributions in $\sqrt{s}$ = 13 TeV pp interactions measured with the ATLAS
  detector at the LHC}'',
  \href{http://dx.doi.org/10.1016/j.physletb.2016.04.050}{{\em Phys. Lett. B}
  {\bfseries 758} (2016) 67--88},
  \href{http://arxiv.org/abs/1602.01633}{{\ttfamily arXiv:1602.01633
  [hep-ex]}}.

\bibitem{CMS:2018nhd}
{\bfseries CMS} Collaboration, A.~M. Sirunyan {\em et~al.}, ``{Measurement of
  charged particle spectra in minimum-bias events from
  proton\textendash{}proton collisions at $\sqrt{s}=13\,\text {TeV} $}'',
  \href{http://dx.doi.org/10.1140/epjc/s10052-018-6144-y}{{\em Eur. Phys. J. C}
  {\bfseries 78} (2018) 697}, \href{http://arxiv.org/abs/1806.11245}{{\ttfamily
  arXiv:1806.11245 [hep-ex]}}.

\bibitem{Aamodt:2008zz}
{\bfseries ALICE} Collaboration, K.~Aamodt {\em et~al.}, ``{The ALICE
  experiment at the CERN LHC}'',
\href{http://dx.doi.org/10.1088/1748-0221/3/08/S08002}{{\em JINST} {\bfseries
  3} (2008) S08002}.

\bibitem{ALICE:2014sbx}
{\bfseries ALICE} Collaboration, B.~Abelev {\em et~al.}, ``{Performance of the
  ALICE Experiment at the CERN LHC}'',
  \href{http://dx.doi.org/10.1142/S0217751X14300440}{{\em Int. J. Mod. Phys. A}
  {\bfseries 29} (2014) 1430044},
  \href{http://arxiv.org/abs/1402.4476}{{\ttfamily arXiv:1402.4476 [nucl-ex]}}.

\bibitem{aliceITS}
{\bfseries ALICE} Collaboration, K.~Aamodt {\em et~al.}, ``{Alignment of the
  ALICE Inner Tracking System with cosmic-ray tracks}'',
  \href{http://dx.doi.org/10.1088/1748-0221/5/03/P03003}{{\em JINST} {\bfseries
  5} (2010) P03003},
\href{http://arxiv.org/abs/1001.0502}{{\ttfamily arXiv:1001.0502
  [physics.ins-det]}}.

\bibitem{Santoro:2009zza}
R.~Santoro {\em et~al.}, ``{The ALICE Silicon Pixel Detector: Readiness for the
  first proton beam}'',
\href{http://dx.doi.org/10.1088/1748-0221/4/03/P03023}{{\em JINST} {\bfseries
  4} (2009) P03023}.

\bibitem{Abelev:2014ffa}
{\bfseries ALICE} Collaboration, B.~Abelev {\em et~al.}, ``{Performance of the
  ALICE Experiment at the CERN LHC}'',
  \href{http://dx.doi.org/10.1142/S0217751X14300440}{{\em Int. J. Mod. Phys.}
  {\bfseries A29} (2014) 1430044},
\href{http://arxiv.org/abs/1402.4476}{{\ttfamily arXiv:1402.4476 [nucl-ex]}}.

\bibitem{forwarddetectorsTdr}
{\bfseries ALICE} Collaboration, P.~Cortese {\em et~al.}, ``{ALICE forward
  detectors: FMD, T0 and V0: Technical Design Report}'', tech. rep., Geneva, 9,
  2004.
\newblock \url{https://cds.cern.ch/record/781854}.

\bibitem{ALICE:2017ban}
{\bfseries ALICE} Collaboration, J.~Adam {\em et~al.}, ``{K$^{*}(892)^{0}$ and
  $\phi(1020)$ meson production at high transverse momentum in pp and Pb-Pb
  collisions at $\sqrt{s_\mathrm{NN}}$ = 2.76 TeV}'',
  \href{http://dx.doi.org/10.1103/PhysRevC.95.064606}{{\em Phys. Rev. C}
  {\bfseries 95} (2017) 064606},
  \href{http://arxiv.org/abs/1702.00555}{{\ttfamily arXiv:1702.00555
  [nucl-ex]}}.

\bibitem{Brun:1994aa}
Brun {\em et~al.}, \href{http://dx.doi.org/10.17181/CERN.MUHF.DMJ1}{{\em
  {GEANT: Detector Description and Simulation Tool; Oct 1994}}}.
\newblock CERN Program Library. CERN, Geneva, 1993.
\newblock \url{https://cds.cern.ch/record/1082634}.
\newblock Long Writeup W5013.

\bibitem{ALICE:2016fzo}
{\bfseries ALICE} Collaboration, J.~Adam {\em et~al.}, ``{Enhanced production
  of multi-strange hadrons in high-multiplicity proton-proton collisions}'',
  \href{http://dx.doi.org/10.1038/nphys4111}{{\em Nature Phys.} {\bfseries 13}
  (2017) 535--539}, \href{http://arxiv.org/abs/1606.07424}{{\ttfamily
  arXiv:1606.07424 [nucl-ex]}}.

\bibitem{Abelev:2012sea}
{\bfseries ALICE} Collaboration, B.~Abelev {\em et~al.}, ``{Measurement of
  inelastic, single- and double-diffraction cross sections in proton--proton
  collisions at the LHC with ALICE}'',
  \href{http://dx.doi.org/10.1140/epjc/s10052-013-2456-0}{{\em Eur. Phys. J. C}
  {\bfseries 73} (2013) 2456}, \href{http://arxiv.org/abs/1208.4968}{{\ttfamily
  arXiv:1208.4968 [hep-ex]}}.

\bibitem{hybridExplanation}
{\bfseries ALICE} Collaboration, B.~Abelev {\em et~al.}, ``{Long-range angular
  correlations on the near and away side in p--Pb collisions at
  $\sqrt{s_\mathrm{NN}}=5.02$ TeV}'',
  \href{http://dx.doi.org/10.1016/j.physletb.2013.01.012}{{\em Phys. Lett.}
  {\bfseries B719} (2013) 29--41},
\href{http://arxiv.org/abs/1212.2001}{{\ttfamily arXiv:1212.2001 [nucl-ex]}}.

\bibitem{UA1:1989bou}
{\bfseries UA1} Collaboration, C.~Albajar {\em et~al.}, ``{A Study of the
  General Characteristics of $p\bar{p}$ Collisions at $\sqrt{s}$ = 0.2 TeV to
  0.9 TeV}'', \href{http://dx.doi.org/10.1016/0550-3213(90)90493-W}{{\em Nucl.
  Phys. B} {\bfseries 335} (1990) 261--287}.

\bibitem{UA5:1986yef}
{\bfseries UA5} Collaboration, G.~J. Alner {\em et~al.}, ``{Scaling of
  Pseudorapidity Distributions at c.m. Energies Up to 0.9 TeV}'',
  \href{http://dx.doi.org/10.1007/BF01410446}{{\em Z. Phys. C} {\bfseries 33}
  (1986) 1--6}.

\bibitem{STAR:2008med}
{\bfseries STAR} Collaboration, B.~I. Abelev {\em et~al.}, ``{Systematic
  Measurements of Identified Particle Spectra in $pp$, d+Au and Au+Au
  Collisions from STAR}'',
  \href{http://dx.doi.org/10.1103/PhysRevC.79.034909}{{\em Phys. Rev. C}
  {\bfseries 79} (2009) 034909},
  \href{http://arxiv.org/abs/0808.2041}{{\ttfamily arXiv:0808.2041 [nucl-ex]}}.

\bibitem{CDF:1989nkn}
{\bfseries CDF} Collaboration, F.~Abe {\em et~al.}, ``{Pseudorapidity
  distributions of charged particles produced in $\bar{p}p$ interactions at
  $\sqrt{s} = 630$ GeV and 1800 GeV}'',
  \href{http://dx.doi.org/10.1103/PhysRevD.41.2330}{{\em Phys. Rev. D}
  {\bfseries 41} (Apr, 1990) 2330--2333}.

\bibitem{CMS:2010tjh}
{\bfseries CMS} Collaboration, V.~Khachatryan {\em et~al.},
  ``{Transverse-momentum and pseudorapidity distributions of charged hadrons in
  $pp$ collisions at $\sqrt{s}=7$ TeV}'',
  \href{http://dx.doi.org/10.1103/PhysRevLett.105.022002}{{\em Phys. Rev.
  Lett.} {\bfseries 105} (2010) 022002},
  \href{http://arxiv.org/abs/1005.3299}{{\ttfamily arXiv:1005.3299 [hep-ex]}}.

\bibitem{CMS:2010wcx}
{\bfseries CMS} Collaboration, V.~Khachatryan {\em et~al.}, ``{Transverse
  Momentum and Pseudorapidity Distributions of Charged Hadrons in pp Collisions
  at $\sqrt{s} = 0.9$ and 2.36 TeV}'',
  \href{http://dx.doi.org/10.1007/JHEP02(2010)041}{{\em JHEP} {\bfseries 02}
  (2010) 041}, \href{http://arxiv.org/abs/1002.0621}{{\ttfamily arXiv:1002.0621
  [hep-ex]}}.

\bibitem{Ames-Bologna-CERN-Dortmund-Heidelberg-Warsaw:1983cqw}
{\bfseries Ames-Bologna-CERN-Dortmund-Heidelberg-Warsaw} Collaboration,
  A.~Breakstone {\em et~al.}, ``{Charged Multiplicity Distribution in p p
  Interactions at ISR Energies}'',
  \href{http://dx.doi.org/10.1103/PhysRevD.30.528}{{\em Phys. Rev. D}
  {\bfseries 30} (1984) 528}.

\bibitem{PHOBOS:2004xnp}
{\bfseries PHOBOS} Collaboration, R.~Nouicer {\em et~al.}, ``{Pseudorapidity
  distributions of charged particles in d + Au and p + p collisions at
  $\sqrt{s_{\rm NN}} = 200$ GeV}'',
  \href{http://dx.doi.org/10.1088/0954-3899/30/8/075}{{\em J. Phys. G}
  {\bfseries 30} (2004) S1133--S1138},
  \href{http://arxiv.org/abs/nucl-ex/0403033}{{\ttfamily
  arXiv:nucl-ex/0403033}}.

\bibitem{ALICE:2012xs}
{\bfseries ALICE} Collaboration, B.~Abelev {\em et~al.}, ``{Pseudorapidity
  density of charged particles in $p$--Pb collisions at $\sqrt{s_{\rm
  NN}}=5.02$ TeV}'',
  \href{http://dx.doi.org/10.1103/PhysRevLett.110.032301}{{\em Phys. Rev.
  Lett.} {\bfseries 110} (2013) 032301},
  \href{http://arxiv.org/abs/1210.3615}{{\ttfamily arXiv:1210.3615 [nucl-ex]}}.

\bibitem{ALICE:2015juo}
{\bfseries ALICE} Collaboration, J.~Adam {\em et~al.}, ``{Centrality dependence
  of the charged-particle multiplicity density at midrapidity in Pb--Pb
  collisions at $\sqrt{s_{\rm NN}}$ = 5.02 TeV}'',
  \href{http://dx.doi.org/10.1103/PhysRevLett.116.222302}{{\em Phys. Rev.
  Lett.} {\bfseries 116} (2016) 222302},
  \href{http://arxiv.org/abs/1512.06104}{{\ttfamily arXiv:1512.06104
  [nucl-ex]}}.

\bibitem{Acharya:2018hhy}
{\bfseries ALICE} Collaboration, S.~Acharya {\em et~al.}, ``{Centrality and
  pseudorapidity dependence of the charged-particle multiplicity density in
  Xe\textendash{}Xe collisions at $\sqrt{s_{\rm NN}}$ =5.44 TeV}'',
  \href{http://dx.doi.org/10.1016/j.physletb.2018.12.048}{{\em Phys. Lett. B}
  {\bfseries 790} (2019) 35--48},
  \href{http://arxiv.org/abs/1805.04432}{{\ttfamily arXiv:1805.04432
  [nucl-ex]}}.

\bibitem{Basu:2020jbk}
S.~Basu, S.~Thakur, T.~K. Nayak, and C.~A. Pruneau, ``{Multiplicity and
  pseudorapidity density distributions of charged particles produced in pp, pA
  and AA collisions at RHIC \& LHC energies}'',
  \href{http://dx.doi.org/10.1088/1361-6471/abc05c}{{\em J. Phys. G} {\bfseries
  48} (2020) 025103}, \href{http://arxiv.org/abs/2008.07802}{{\ttfamily
  arXiv:2008.07802 [nucl-ex]}}.

\end{thebibliography}\endgroup
